\documentclass[onecolumn,showpacs,preprintnumbers,amsmath,amssymb]{revtex4}

\usepackage{graphicx}
\usepackage{dcolumn}
\usepackage{bm}

\begin{document}

\title{ $U_L(2)\bigotimes U_R(2) $  Model of  Electro-Weak Interaction }

\author{A.V.Koshelkin.}
 \altaffiliation[$ A\underline ~ Kosh@internets.ru; koshelkin@mtu-net.ru;$ ]
 {}
\affiliation{Moscow Institute for Physics and Engineering,
Kashirskoye sh., 31, 115409 Moscow, Russia }
\date{\today}

\begin{abstract}
The $U_L(2)\bigotimes U_R(2) $ gauge model for the unified theory
of the electromagnetic and weak interactions which is free from a
prior self-interaction scalar field,  is developed. Due to
breaking the initial symmetry  the $SU_L(2)\bigotimes U_R(1) $
Lagrangian is derived. The obtained $SU_L(2)\bigotimes U_R(1) $
Lagrangian contains the  whole of terms corresponding  both to
free boson and fermion fields and to interaction between them, as
it takes place in the Standard Model (SM) .  We show that all
boson fields, including the Higgs one, directly arise due to
breaking the initial symmetry,  and are  generated by the initial
gauge fields in contrary to the Standard Model consideration. The
Higgs fields are studied in detail. A broad spectrum of states of
the Higgs bosons is found. The masses of the Higgs particle in
such states are calculated.

\end{abstract}


\maketitle

\section{Introduction}

The first attempt to unify the electromagnetism and weak
interaction have been made by H.Yukawa\cite{1}  and
O.Klein\cite{2} where the boson-exchange model for the
charge-changing weak interactions was proposed. J.Schwinger has
studied this problem in the 1950s\cite{3}. The hitch  in the
unification of the electromagnetic and weak interaction has been
made due to the papers by S.Glashow\cite{4},S.Weinberg\cite{5}, A.
Salam\cite{6}. The developed  Gleshow-Weiberg-Salam theory
appeared to be in a good agree with the   neutral currents
experiments\cite{7,8,9}. Since the papers\cite {4,5,6} have been
issued the unified theory of weak and electromagnetic interactions
is one of the most important and discussable problem in the
physics of particles and  fields.

The observations of $W^\pm$ and $Z^{(0)}$ bosons by the UA1 and
UA2 Collaborations in CERN in 1983\cite{10}, which supported the
Glashow-Weinberg-Salam model, have  additionally  increased
interest to the Standard Model of electro-weak interaction due to
the problem of the  Higgs boson\cite{11} which has been still
undiscovered.

Searches  for the Higgs boson  in the Standard Model  and beyond
it have still  remained  the main goals  in investigations of both
the Tevatron (CDF and D0 Colloborations) and  LHC (ATLAS and CMS
Colloborations). In this way, the present LEP, Tevatron and LHC
data have already placed strong direct bounds on the possible mass
of the  Higgs particle. According to the result obtained them the
allowed range of the Higgs mass is between 114 GeV and 145 GeV,
and above 450 GeV \cite{12, 13,14,15}.

Studying the Higgs boson strongly  stimulates developments of
various theoretical  models beyond  SM which can be, generally,
verified due to the Tevatron and LHC machines. The LHC results
have been recently considered in the context of SUSY singlet
extensions \cite{16}, in more general 2HDM scenarios \cite{17},
and also for dark Higgs models \cite{18} (where the SM Higgs
sector is enlarged with the SM singlet). In the past years there
has been extensive work in extensions of the Higgs sector in terms
of the Minimal Supersymmetric Standard Model (MSSM) by
higher-dimension operators \cite{19} and due to models beyond MSSM
(BMSSM)\cite{20}.

In the present paper  unification of the electromagnetic and weak
interaction is studied in terms of the $U_L(2)\bigotimes U_R(2) $
gauge symmetry without a priory self-interaction scalar field. Due
to braking the initial symmetry the $SU_L(2)\bigotimes U_R(1) $
Lagrangian is derived. The obtained $SU_L(2)\bigotimes U_R(1) $
Lagrangian is found to take into account correctly of both the
charged and neutral interaction  currents as it takes place in the
SM\cite{4,5,6}. In this way, all massive boson fields (including
the Higgs one), the  massive fermion field as well as
electromagnetic field naturally arise as the superposition of
various modes of the initial  gauge fields without any additional
self-interacting scalar field as compared with the situation
taking place in the SM consideration\cite{4,5,6}. The structure of
the Higgs field is studied in detail. On a basis of the
experimental data for the fine structure coupling constatnt, Fermi
coupling constants,  and $W$ boson mass, the coupling constant for
the interaction current as well as the masses of the Higgs and
$Z^{(0)}$ bosons  are calculated.

It is shown that there is a broad spectrum of the mass states of
the Higgs boson which are formed by various modes of the initial
gauge fields. The masses of such states are calculated. The Higgs
boson mass is found to be in the interval $M_H = 92.8 \div 231.7
~GeV$.

The paper is organized as follows. The second section contains the
statement of the problem. Breaking the initial symmetry and
deriving the $SU_L(2)\bigotimes U_R(1) $ Lagrangian are in Section
III. The qualitative calculations of the coupling constants and
masses are in the  section IV. The last section is Conclusion.

\section{Statement of the problem }

We start from the $U_L (2)\bigotimes U_R (2)$ gauge invariant
Lagrangian:

\begin{eqnarray}
&& {\cal L} =  {\cal L}_{(e)} + {\cal L}_{(\mu)} + {\cal
L}_{(\tau)};
\nonumber \\
 && {\cal L}_{(e)}  =   Tr ~ \Bigg\{
{i\over 2}  \left( {\bar \Psi}^{(e)} (x) \gamma^\mu
\partial_\mu  \  \Psi^{(e)} (x)  -   {\bar \Psi}^{(e)} (x)
\gamma^\mu {\overleftarrow \partial_\mu } \ \Psi^{(e)} (x) \right)
+ g_1 ~{\bar \Psi}^{(e)}_L (x) \gamma^\mu {\cal A}^a_\mu (x) ~ T_a
~\Psi^{(e)}_L (x) -  \nonumber \\
&& g_2 ~{\bar \Psi}^{(e)}_R (x) \gamma^\mu {\cal B}^a_\mu (x)~ T_a
~\Psi^{(e)}_R (x) \Bigg\} ~ - ~ {1\over 4 } \left[ F^a_{\mu \nu}
(x) \ F_a^{\mu \nu} (x) + G^a_{\mu \nu} (x) \ G_a^{\mu \nu} (x)
\right] \equiv {\cal L}^{(free)}_{(e)} - {\cal L}^{(int)}_{(e)} ,
\end{eqnarray}
where the indexes $(e )$, $(\mu )$, $(\tau )$ correspond to the
electronic, muonic and $\tau$-leptonic parts of the Lagrangian.
The Lagrangians ${\cal L}_{(\mu)}$ and $ {\cal L}_{(\tau)}$ are
obtained  from ${\cal L}_{(e)}$ due to  the  substitutions of the
index $(e)$ by $(\mu)$ and $(\tau)$, respectively\footnote{We
study the electronic part of the Lagrangian (1) since all results
derived below are directly extended to the other terms of it by
means of changing  corresponding notations. }. The superscripts
mean the part of the Lagrangian which correspond to free particles
and interactions between them, respectively.

The "left"  $\Psi^{(e)}_L (x)$ and "right"  $\Psi^{(e)}_R (x)$
doublets of the   massless fermion field $\Psi^{(e)} (x)$ are
given by the formulae:

\begin{eqnarray}
&&\Psi^{(e)}_L (x) = \frac{1 + \gamma^5}{2}~~ \Psi^{(e)} (x) ; ~~~
\Psi^{(e)}_L (x) = \frac {1 - \gamma^5}{2}~~ \Psi (x) ; ~~~
\Psi^{(e)} (x) = \left(
\begin{array}{cccc} \nu^{(e)} (x)
  \\
e(x)
  \end{array} \right),
\end{eqnarray}
where $\nu^{(e)} (x) ~; ~ e (x)$ are   neutrino and massless
electron fields, respectively; $\gamma^k$ are the Dirac matrixes.

The tensors of the gauge fields are determined by the standard
way\cite{21}:

\begin{eqnarray}
&& F^a_{\mu \nu}(x) =
\partial_\mu {\cal A}^a_\nu (x) - \partial_\nu {\cal A}^a_\mu (x) + g_1 ~ f^a_{~~bc} ~ {\cal A}^b_\mu (x) ~  {\cal
A}^c_\nu (x) \equiv {\cal A}^a_{\mu  \nu} (x) ~ {\cal A}_a^{\mu
\nu} (x) + g_1 ~ f^a_{~~bc} ~ {\cal A}^b_\mu (x) ~  {\cal A}^c_\nu
(x) ; \nonumber \\  \nonumber \\
&& G^a_{\mu \nu} (x) =
\partial_\mu {\cal B}^a_\nu (x) - \partial_\nu {\cal B}^a_\mu  (x) +  g_2 ~ f^a_{~~bc} ~{\cal B}^b_\mu (x) ~  {\cal
B}^c_\nu(x)  \equiv {\cal B}^a_{\mu  \nu} (x) ~ {\cal B}_a^{\mu
\nu} (x) + g_1 ~ f^a_{~~bc} ~ {\cal B}^b_\mu (x) ~ {\cal B}^c_\nu
(x)
\end{eqnarray}
where   $g_1$ and $g_2$ are the coupling constants. The symbols
$f^a_{~~bc}$ are the structure of constants of the $U(2)$ group
which govern the commutative relations between the generators
$T_a$:

\begin{eqnarray}
&& T_0  = \left( \begin{array}{cccc} 1 \
\ \ \ \ \  0  \\ \\
0 \ \ \ \ \ \  1  \\
 \end{array} \right) ; \ \ \ T_1   = \left( \begin{array}{cccc} 0 \
\ \ \ \ \  1   \\ \\
1 \ \ \ \ \ \  0
 \end{array} \right); T_2  =\left( \begin{array}{cccc} 0 \
\ \ \ \ \  - i   \\ \\
i \ \ \ \ \ \  0  \\
 \end{array} \right) ;  T_3 = \left( \begin{array}{cccc} 1 \
\ \ \ \ \  0  \\ \\
0 \ \ \ \ \ \  - 1  \\
 \end{array} \right) ; \nonumber \\ \nonumber \\
&&  \left[ T_a, T_b \right]_-    =  T_a  T_b - T_b T_a =   i
f_{ab}^{\ \ c} T_c ; ~~~  Tr \ ( T_a \ T_b )   ={1\over 2}
\delta_{ab} \left( 1 + \frac{3(\delta_{a0} + \delta_{b0} )}{2}
\right)
\end{eqnarray}

The constant $f^a_{~~bc}$ are

\begin{eqnarray}
&& f_{abc} = \pm 1 ;~~ when ~~ a\ne b \ne c ; ~~  a,b,c \ne 0; ~~~
f_{123} = + 1 .
\end{eqnarray}
In this  way,  $f^a_{~~bc} =0$ , when either any two indexes are
the same,  or some  index is equal to zero.

In the formulae (1)-(4) we take that  $x\equiv x^\mu = (x^0 ;
{\vec x})$  is a vector in the Minkowski space-time;
$\partial_{\mu} = (\partial /\partial t ; \nabla )$.  We use the
signature $ diag \left( {g}^{\mu \nu} \right) = (1; -1; -1; -1)$
for the metric tensor ${g }^{\mu \nu}$. The line   over $\Psi$
means the  Dirac   conjugation.

\section {Breaking the initial symmetry}

We break the $U(2)\bigotimes U(2)$ symmetry by shifting the gauge
fields ${\cal A}^a_\nu (x)$ and ${\cal B}^a_\nu (x)$ according to
the formula:

\begin{eqnarray}
&&{\cal A}^a_\nu (x) ={ \textmd a}^a_\nu  + { A}^a_\nu (x) ; ~~~
{\cal B}^a_\nu (x) = {\textmd b}^a_\nu  + { B}^a_\nu (x);
\end{eqnarray}
where ${ \textmd a}^a_\nu$ and ${ \textmd b}^a_\nu$ are the
constant vectors such that

\begin{eqnarray}
&&{ \textmd a}^b_\nu~{ \textmd a}_c^\mu = -
\frac{1}{16}{g}_\nu^\mu~\delta^b_c ~ { \textmd a}^2 ;~~~ { \textmd
b}^a_\nu~{ \textmd b}_c^\mu = - \frac{1}{16}{g}_\nu^\mu~\delta^a_c
~ { \textmd a}^2 , ~~~ a,b = 0,1,2,3;
\end{eqnarray}
where ${ \textmd a}$ and ${ \textmd b}$ are  some real constants.

We also simultaneously change  the "left" and "right" components
of the  fermion fields  by the following unitarian transformation
in the Dirac space:

\begin{eqnarray}
&&\Psi^{(e)}_L (x)=  \left\{(T_{l(X;x)} \exp \right\} \left\{ i
g_1 a^a_\mu T_a (x^\mu - X^\mu ) + i g_1 T_3 \int\limits_{X}^x
dx^\mu A_\mu^3 + ig_1 T_3 \int\limits_{X}^x dx^\mu A^0_\mu + \frac
{i g_1}{2} (3~ T_0 - T_3 ) \int\limits_{X}^x dx^\mu B_\mu^3
\right\} ~ \cdot
\nonumber \\
&& \left\{(T_{l(X;x)} \exp \right\} \left\{ i \frac{g_1} {2}
\int\limits_{X}^x dx^\mu \left( T_1 A^1_\mu + T_2 A^2_\mu \right)
+ i \frac{ g_1 g_2}{2(g^2_1 + g^2_2 )} (T_3 -T_0 )
\int\limits_{X}^x dx^\mu \left( g_1 B^0_\mu - g_2 A^0_\mu \right)
\right\} \cdot
 \nonumber \\
&& \left\{(T_{l(X;x)} \exp \right\} \left\{  i\frac{ G_{(e)}}{4}
\gamma_\mu (T_0 - T_3 ) \int\limits_{X}^x dx^\mu \left( B^1_\nu
\epsilon_1^\nu   + B^2_\nu \epsilon_2^\nu + A^3_\nu \epsilon_3^\nu
\right) \right\}~\psi^{(e)}_L (x)
\end{eqnarray}

\begin{eqnarray}
&&\Psi^{(e)}_R (x)=  \left\{(T_{l(X;x)} \exp \right\} \left\{- i
g_2 b^a_\mu T_a (x^\mu - X^\mu ) - i g_2 \int\limits_{X}^x dx^\mu
\sum\limits_{a=1}^3 ~ T_a~ B^a_\mu -  i g_2 T_3 \int\limits_{X}^x
dx^\mu B_\mu^0 \right\} ~ \cdot \nonumber \\
&& \left\{(T_{l(X;x)} \exp \right\} \left\{ i \frac{ g_1
g_2}{2(g^2_1 + g^2_2 )} (T_3 -T_0 ) \int\limits_{X}^x dx^\mu
\left( g_1 B^0_\mu -
g_2 A^0_\mu \right)\right\} ~ \cdot \nonumber \\
&& \left\{(T_{l(X;x)} \exp \right\} \left\{ i\frac{ G_{(e)}}{4}
\gamma_\mu (T_0 - T_3 ) \int\limits_{X}^x dx^\mu \left( B^1_\nu
\epsilon_1^\nu   + B^2_\nu \epsilon_2^\nu + A^3_\nu \epsilon_3^\nu
\right)  \right\} ~\psi^{(e)}_R (x) ,
\end{eqnarray}
where $G_{(e)}$ is some coupling  constant. The symbol
$\left\{(T_{l(X;x)} \exp \right\}$ is the chronological exponent
which means integration along the line $l(X;x)$ in the Minkowski
space-time so that the points $X$ is always before the point $x$.

The parameters ${ \textmd a}$ and { \textmd b} introduced in
Eqs.(6) and (7) as well as the coupling constant $G_{(e)}$ can be
expressed in terms of the observable masses and coupling constants
as it will be shown below (see Discussion).

We substitute the $\Psi^{(e)}_L (x)$ and $\Psi^{(e)}_R (x)$
transformed according to Eqs.(8), (9) into the Lagrangian (1).
After that, following the Schwinger's idea \cite{22} we calculate
a limit:

\begin{eqnarray}
&& {\cal L} [{\bar \Psi} (x); ~~ \Psi (x)] = \lim \limits_{x' \to~
x} ~~ {\cal L} [{\bar \Psi} (x'); ~~ \Psi (x)].
\end{eqnarray}

As a result, the kinematic term  of the fermion part of the
Lagrangian (1) takes the form:

\begin{eqnarray}
&&   Tr ~  {i\over 2}  \left( {\bar \Psi}^{(e)} (x) \gamma^\mu
\partial_\mu    \Psi^{(e)} (x)  -   {\bar \Psi}^{(e)} (x)
\gamma^\mu {\overleftarrow \partial_\mu }  \Psi^{(e)} (x) \right)
=   Tr ~  {i\over 2}  \left( {\bar \psi}^{(e)} (x) \gamma^\mu
\partial_\mu    \psi^{(e)} (x)  -   {\bar \psi}^{(e)} (x)
\gamma^\mu {\overleftarrow \partial_\mu }  \psi^{(e)} (x) \right)
+ \nonumber \\
&&\lim \limits_{x' \to~ x} \Bigg\{ Tr ~ i \Bigg[  {\bar
\psi}_L^{(e)} (x) \gamma^\mu
\partial_\mu \Bigg\{ \exp  \Bigg( i g_1
a^a_\mu T_a ( x^\mu - x'^\mu )  + i g_1 T_3 \int\limits_{x'}^x
dx^\mu A_\mu^3 +  \frac {i g_1}{2} (3~T_0 - T_3 )
\int\limits_{x'}^x dx^\mu B_\mu^3 + ig_1 T_3 \int\limits_{x'}^x
dx^\mu A^0_\mu \Bigg) \cdot \nonumber
\\
&& \exp \left( i \frac{g_1} {2} \int\limits_{x'}^x dx^\mu \left(
T_1 A^1_\mu + T_2 A^2_\mu \right) + i \frac{ g_1 g_2}{2(g^2_1 +
g^2_2 )} (T_3 -T_0 ) \int\limits_{x'}^x dx^\mu \left( g_1 B^0_\mu
- g_2 A^0_\mu \right)  \right) \cdot
 \nonumber \\
&& \exp \left( i\frac{ G_{(e)}}{4} \gamma_\mu (T_0 - T_3 )
\int\limits_{x'}^x dx^\mu \left( B^1_\nu \epsilon_1^\nu   +
B^2_\nu \epsilon_2^\nu + A^3_\nu \epsilon_3^\nu \right)   \right)
\Bigg\} { \psi}_L^{(e)} (x)   \Bigg] + \nonumber
\end{eqnarray}

\begin{eqnarray}
 && Tr ~ i \Bigg[  {\bar \psi}_R^{(e)} (x) \gamma^\mu
\partial_\mu \Bigg\{ \exp \left(- i g_2
b^a_\mu T_a ( x^\mu - x'^\mu ) - i g_2 \int\limits_{x'}^x dx^\mu
\sum\limits_{a=1}^3 ~ T_a~ B^a_\mu -  i g_2 T_3 \int\limits_{x'}^x
dx^\mu B_\mu^0 \right) ~ \cdot \nonumber \\
&& \exp \left( i\frac{ G_{(e)}}{4} \gamma_\mu (T_0 - T_3 )
\int\limits_{x'}^x dx^\mu \left( B^1_\nu \epsilon_1^\nu   +
B^2_\nu \epsilon_2^\nu + A^3_\nu \epsilon_3^\nu \right) + i \frac{
g_1 g_2}{2(g^2_1 + g^2_2 )} (T_3 -T_0 ) \int\limits_{x'}^x dx^\mu
\left( g_1 B^0_\mu - g_2 A^0_\mu \right) \right) \Bigg\}
 {\psi}_R^{(e)} (x)   \Bigg] \Bigg\}, \nonumber \\
\end{eqnarray}
where the differentiation $\partial_\mu $ with respect to the
unprimed variable $x$ is assumed in the last formula. The symbols
$\epsilon^a_\nu$ are the constant vectors which is governed by the
expressions:

\begin{eqnarray}
&& \epsilon^a_\nu = e^a_\nu + e'^a_\nu ; ~~~~  e^a_\nu e_b^\mu = -
\frac{1}{12}  {\cal G}^\mu_\nu ~\delta^a_b ; ~~e'^a_\nu e'^\mu_b =
- \frac{1}{12} {\cal G}^\mu_\nu ~\delta^a_b ;~~ e'^a_\mu e_b^\mu =
0; ~~~ a,b = 1,2,3.
\end{eqnarray}

We calculate the limit in Eq.(11) when $x'$ goes to $x$. After
that, substituting the transformations (6), (8), (9) into the the
initial Lagrangian we derive:

\begin{eqnarray}
&& {\cal L}^{(free)}_{(e)} = Tr ~  {i\over 2}  \left( {\bar
\psi}^{(e)} (x) \gamma^\mu
\partial_\mu    \psi^{(e)} (x)  -   {\bar \psi}^{(e)} (x)
\gamma^\mu {\overleftarrow \partial_\mu }  \psi^{(e)} (x) \right)
- \nonumber \\
&&  {1\over 4 } \left\{ A^a_{\mu \nu} (x) \ A_a^{\mu \nu} (x) +
B^a_{\mu \nu} (x) \ B_a^{\mu \nu} (x) - \frac{3}{4}~ g^2_1 {
\textmd a}^2 A^a_{\mu} (x) \ A_a^{\mu} (x) - \frac{3}{4}~  g_2^2 {
\textmd b}^2 B^a_{\mu} (x) \ B_a^{\mu} (x) \right\} + {\cal
L}_{(e)} [A^4 ; B^4 ]
\end{eqnarray}

\begin{eqnarray}
&& - {\cal L}^{(int)}_{(e)} =  Tr ~   \Bigg\{\frac {g_1}{2} {\bar
\psi}_L^{(e)} (x) \gamma^\mu ( T_1 A^1_\mu + T_2 A^2_\mu )
\psi_L^{(e)} (x) + g_1 {\bar \psi}_L^{(e)} (x) \gamma^\mu ( T_0 -
T_3 ) A^0_\mu \psi_L^{(e)} (x) - \nonumber \\
&&  \frac{g_1}{2} {\bar \psi}_L^{(e)} (x) \gamma^\mu (3~ T_0 -T_3
)~ B^3_\mu \psi_L^{(e)} (x) +  g_2  {\bar \psi}_R^{(e)} (x)
\gamma^\mu ( T_3 - T_0 ) B^0_\mu \psi_R^{(e)} (x)
 \Bigg\} - \nonumber \\
 &&  Tr ~ \Bigg\{ \frac{G_e}{2}~  {\bar e } (x) ~
 \left( B^1_\nu \epsilon_1^\nu   + B^2_\nu \epsilon_2^\nu +
A^3_\nu \epsilon_3^\nu \right) ~ e (x) + \frac{g_1 g_2 }{ (g_1^2 +
g_2^2 )} {\bar e } (x) ~ \gamma^\mu( g_1 B^0_\mu - g_2 A^0_\mu ) ~
e (x) \Bigg\},
\end{eqnarray}
where ${\cal L}^{(free)}_{(e)}$,  ${\cal L}^{(int)}_{(e)}$ are the
Lagrangians  of free field  and interaction between them,
respectively; ${\cal L}_{(e)} [A^4 ; B^4 ]$ is the part of the
Lagrangian which contains the gauge field in the 4th power. We
keep the old notations $e (x)$ for the electronic field in
Eq.(14), although the transformations (9), (10) have been already
made.

Following the standard way we introduce the charged boson  fields
$W^\pm_\nu (x) $\cite{5,6} as well as the electromagnetic $A_\nu
(x) $ and neutral boson field $Z^{(0)}_\nu (x)$:

\begin{eqnarray}
&& A^1_\nu = \frac {W^-_\nu + W_\nu^+ }{\sqrt2} ; ~~ A^2_\nu =
\frac {W^-_\nu - W_\nu^+ }{i \sqrt2};~~\nonumber \\
&& A^0_\nu = \frac{- g_2 A_\nu + g_1 Z_\nu}{\sqrt{g^2_1 + g^2_2}};
~~ B^0_\nu = \frac{ g_1 A_\nu + g_2 Z_\nu}{\sqrt{g^2_1 +
g^2_2}};~~ B^3_\nu = \frac{\sqrt{g^2_1 + g^2_2}}{2 g_1} ~ Z_\nu;
~~ Z_\nu = \left(  \frac{4 g_1^2}{5 g_1^2 + g_2^2
}\right)^{\frac{1}{2}} ~ Z^{(0)}_\nu
\end{eqnarray}

As to the components $B_\nu^1$, $B_\nu^2$ and $A_\nu^3$ we
determine them by means of the realtions:

\begin{eqnarray}
\ && B_\nu^1 =  (\sigma_{(1)} (x) e_\nu^1 + \sigma_0 e'^1_\nu );
~~ B_\nu^2 =  (\sigma_{(2)} (x) e_\nu^2 + \sigma_0 e'^2_\nu ); ~~
A_\nu^3 =  (\sigma_{(3)} (x) e_\nu^3 + \sigma_0 e'^3_\nu ),
\end{eqnarray}
where $\sigma_{(i)}$ are scalar functions while the constant
$\sigma_0$ is related to the electron mass $m_e$ and the coupling
constant $G_e$ (see Eqs.(8), (9)) so that:

\begin{eqnarray}
m_e = - \frac{G_e \sigma_0}{2}.
\end{eqnarray}

Substituting $A_\nu^a$ and $B_\nu^a$ in the form given by
Eqs.(15)-(17) into the formulae (13), (14) we derive:

\begin{eqnarray}
&& {\cal L}^{(free)}_{(e)} =   {i\over 2}  \left( {\bar e} (x)
\gamma^\mu \partial_\mu    {e} (x)  -   {\bar e} (x) \gamma^\mu
{\overleftarrow \partial_\mu }  e (x) \right) + {i\over 2}  \left(
{\bar \nu}^{(e)} (x) \gamma^\mu \partial_\mu    {\nu}^{(e)} (x)  -
{\bar \nu}^{(e)} (x) \gamma^\mu {\overleftarrow \partial_\mu }
\nu^{(e)} (x) \right) - m_e ~ {\bar e} (x) ~  {e} (x)
- \nonumber \\
&&  - \frac{1}{2} {W^-}^{\mu \nu} {W^+}_{\mu \nu} + \frac{3 g_1^2
a^2}{8} ~ {W^-}^{\mu } {W^+}_{\mu } - \frac{1}{4} {Z^{(0)}}^{\mu
\nu} {Z^{(0)}}_{\mu \nu} + \frac{ 3 g_2^2 b^2}{16} ~ \frac{  g_1^2
+ g_2^2 }{5 g^2_1 + g^2_2} ~  {Z^{(0)}}^{\mu } {Z^{(0)}}_{\mu }
-\frac{1}{4}
{A}^{\mu \nu} {A}_{\mu \nu} + \nonumber \\
&&  \frac{1}{8} \sum\limits_{1}^3  (\partial_\mu \sigma_{(i)} (x))
(\partial^\mu \sigma_{(i)} (x)) - \frac{1}{16}\left( g_1^2 a^2 ~~
\sigma_{(3)} (x) + g_2^2 b^2 ~  \left( \sigma_{(1)} (x) +  ~
\sigma_{(2)} (x)  \right) \right) +{\cal L}_{(e)} [A^4 ; B^4 ]
\end{eqnarray}

\begin{eqnarray}
&& - {\cal L}^{(int)}_{(e)} =   {g_1\over 2 \sqrt{2}}  \left(
{\bar \nu}^{(e)} (x) \gamma^\mu (1+\gamma^5 ) ~ {e} (x) ~ W^+_\mu
+ {\bar e} (x) \gamma^\mu (1+\gamma^5 )  \nu^{(e)} (x) ~ W^-_\mu
\right) - \frac{ g_1 g_2 }{\sqrt{g^2_1 + g^2_2}} ~{\bar e} (x)
\gamma^\mu {e} (x)
A_\mu - \nonumber \\
&&  \frac{ g_1 \sqrt{g^2_1 + g^2_2 }  }{2 \sqrt{5 g^2_1 + g^2_2}}
\left\{ ~{\bar \nu}^{(e)} (x) \gamma^\mu \left(  1  + \gamma^5
\right) \nu^{(e)} (x) Z^{(0)}_\mu - ~ 2~ {\bar e} (x) \gamma^\mu
\left( \frac{ g^2_1 - 3 g^2_2   }{g^2_1 + g^2_2} + \gamma^5
\right) {e} (x) Z^{(0)}_\mu \right\} -
\nonumber \\
&&\frac{G_e }{6} ~\sum\limits_{1}^3  {\bar e} (x) \sigma_{(i)} (x)
{e} (x).
\end{eqnarray}
In the last formula we  kept the old notations for the neutrino
$\nu^{(e)} (x)$ and electron $e(x)$ fields although they have been
already transformed by means of Eqs.(8), (9).

 Let us introduce the masses of the intermediate charged
$W^\pm$ and neutral $Z^{(0)}$ bosons according to the formulae:

\begin{eqnarray}
&&  M_W^2 = \frac{3}{8} g_1^2 { \textmd a}^2~; ~~~  M_Z^2 =
\frac{3}{8} g_2^2 { \textmd b}^2 ~\frac{  g_1^2 + g_2^2}{5 g^2_1 +
g^2_2}; ~~ { \textmd b}^2 = { \textmd a}^2 ~ \frac{ 5 g_1^2 +
g_2^2}{ g_2^2}.
\end{eqnarray}

It follows from  Eqs.(17), (20) that the parameters ${ \textmd
a}$, ${ \textmd b}$ and  $G_e$ are expressed via the observable
coupling constants and masses as it has been already mentioned
before.

Then, the  Lagrangian of free  particles can be written as
follows:

\begin{eqnarray}
&& {\cal L}^{(free)}_{(e)} =   {i\over 2}  \left( {\bar e} (x)
\gamma^\mu \partial_\mu    {e} (x)  -   {\bar e} (x) \gamma^\mu
{\overleftarrow \partial_\mu }  e (x) \right) + {i\over 2}  \left(
{\bar \nu}^{(e)} (x) \gamma^\mu \partial_\mu    {\nu}^{(e)} (x)  -
{\bar \nu}^{(e)} (x) \gamma^\mu {\overleftarrow \partial_\mu }
\nu^{(e)} (x) \right) - m_e ~ {\bar e} (x) ~  {e} (x)
- \nonumber \\
&&  - \frac{1}{2} {W^-}^{\mu \nu} {W^+}_{\mu \nu} + M_W^2  ~
{W^-}^{\mu } {W^+}_{\mu } - \frac{1}{4} {Z^{(0)}}^{\mu \nu}q
{Z^{(0)}}_{\mu \nu} + \frac{1}{2} ~ M_Z^2 ~  {Z^{(0)}}^{\mu }
{Z^{(0)}}_{\mu } -\frac{1}{4}
{A}^{\mu \nu} {A}_{\mu \nu} + \nonumber \\
&&  \frac{1}{8} \sum\limits_{1}^3  (\partial_\mu \sigma_{(i)} (x))
(\partial^\mu \sigma_{(i)} (x)) - \frac{1}{6}\left(  M_W^2  ~~
\sigma^2_{(3)} (x) + \left(  5 + \frac{ g_2^2}{ g^2_1 } \right)~
M_W^2 \left[ \sigma^2_{(1)} (x) + ~ \sigma^2_{(2)} (x) \right]
\right) +{\cal
L}_{(e)} [A^4 ; B^4 ] \equiv \nonumber \\
&&{\cal L}^{(free)}_{(e)} (lepton) + {\cal L}^{(free)}_{(e)} (W) +
{\cal L}^{(free)}_{(e)} (Z) + {\cal L}^{(free)}_{(e)} (Higgs)
+{\cal L}_{(e)} [A^4 ; B^4 ],
\end{eqnarray}
where $W^\pm_\mu$ and $Z^{(0)}_\mu$ are the massive vector fields,
$A_\mu$ is the electromagnetic field. The functions $\sigma_{(i)}
(x)$ form the field of a scalar boson.

Eqs. (19), (21) consist of the electro weak Lagrangian which is
$SU_L(2)\bigotimes U_R(1)$ invariant. Although the Lagrangian
given by Eqs.(19), (21) is very similar in its structure to the
one derived by S.Weinberg and A.Salam\cite{5,6} , it dramatically
differ from the Weinberg-Salam Lagrangian since all terms in the
formulae (18), (19) are governed by the initial gauge fields
without any additional scalar field interacting with fermion and
gauge fields\cite{5,6}.

\section {Discussion}

Following the standard way we take

\begin{eqnarray}
&&  \sqrt{\alpha} = \frac{g_1 g_2}{\sqrt {g_1^2 + g_2^2}}:~
\frac{G_F}{\sqrt{2}} = \frac{g_1^2}{8 M^2_W}; ~ \sin \theta_W =
\frac{\sqrt{\alpha}}{g_1} = \frac{g_2 }{\sqrt {g_1^2 + g_2^2}},
\end{eqnarray}
where $\theta_W$ is the Weinberg angle;  $\alpha$ and $G_F$ are
the fine structure coupling constant and the Fermi coupling
constants, respectively. Provided that $M_W$ and $\theta_W$ are
known due to experiments, the formulae (20), (22) establish
relation between $M_W$ and $M_Z$:

\begin{eqnarray}
&&  M^2_Z = \frac{ M^2_W} {\cos^2 \theta_W}.
\end{eqnarray}

Setting $\sin^2 \theta_W = 0.23 $\cite{22} we obtain the
well-known result for the $Z^{(0)}$ boson mass\cite{23}.

Taking into account of the experimental results \cite{23}, we
obtain that the constants $g_1$ and $g_2$ are:

\begin{eqnarray}
g_1 \thickapprox 0.65 ,~~ g_2 \thickapprox 0.36
\end{eqnarray}

\subsection{Interaction Lagrangian }

Comparing the interaction Lagrangian (19) with the corresponding
terms in the  Standard Model calculations\cite{5,6} we have found
that the Lagrangians  slightly differ by the coupling  constant
$C(neutral )$ for a neutral current. They are connected  by the
relation:

\begin{eqnarray}
&&C(neutral) = C_{SM}(neutral) \cdot \frac{2 }{\sqrt{ 5 + g_2^2 /
g_1^2}} \simeq 0.87 ~ C_{SM}(neutral),
\end{eqnarray}
where $C_{SM} (neutral)$ is the coupling  constants corresponding
to the neutral  currents in the Standard Mode\cite{5,6}.

\subsection{Higgs bosons}

The terms in Eq.(21)  which depend on the scalar functions
$\sigma_{(i)} (x)$ contain  the main differences from the SM
results. First, they as well as the whole of the Lagrangian (19),
(21) are derived in terms of the gauge fields ${\cal A}_\mu^a $
and ${\cal B}_\mu^a $ without any additional scalar
field\cite{5,6}. Besides that, various superpositions of
$\sigma_{(i)} (x)$ appear to  form the different mass states of
the scalar Higgs  mode. The structure of the Lagrangian (21)
dictates the following mass states to the Higgs boson. It is
obvious that there are possible  either a singlet or doublet or
singlet and doublet together. Let us study  mass states  of the
scalar Higgs mode.

 a) When the all fields $\sigma_{(i)} (x)$ are different and
not equal to zero, so that $\sigma_{(1)} \ne \sigma_{(2)} \ne
\sigma_{(3)}$, the Lagrangian ${\cal L}^{(free)}_{(e)} (Higgs)$
can be presented in the form:

\begin{eqnarray}
&& {\cal L}^{(free)}_{(e)} (Higgs) =  \frac{1}{2}(\partial_\mu H_S
) (\partial^\mu H_S ) - \frac{1}{2} M^2_S (H) ~H^2_S
  + \nonumber \\
&& \frac{1}{2}~
\partial_\mu
\left( \begin{array}{cccc} H_D (1)  \\ \\
H_D (2)  \\
 \end{array} \right)^\dag \cdot \partial^\mu
\left( \begin{array}{cccc} H_D (1)  \\ \\
  H_D (2) \\
 \end{array} \right) - \frac{1}{2}
 ~ M^2_D (H) \left( \begin{array}{cccc} H_D (1)  \\ \\
H_D (2)  \\
 \end{array} \right)^\dag \cdot \left( \begin{array}{cccc} H_D (1)  \\ \\
  H_D (2) \\
 \end{array} \right) \nonumber \\ \nonumber \\
&& H_S = \frac{\sigma_{(3)} (x)}{{2}} ;~~ H_D (1) =
\frac{\sigma_{(1)} (x)}{2}: ~ H_D (2)  = \frac{\sigma_{(2)}
(x)}{2}
\end{eqnarray}
where $M_D (H) $ and $M_S (H) $ are the masses of a  doublet and
singlet, respectively, which are

\begin{eqnarray}
&& M_D (H)= M_W \sqrt{ \frac{20}{3} + \frac{4 g_2^2}{3 g^2_1 }}~
 \thickapprox 213.7 GeV ; ~~~~~~
  M_S (H) = M_W \sqrt{ \frac{4}{3}} \thickapprox 92.8
 GeV.
\end{eqnarray}

b) If $\sigma_{(3)} (x) = 0 $ but $\sigma_{(1)} (x) \ne
\sigma_{(2)} (x)$ the Lagrangian has the form:

\begin{eqnarray}
&& {\cal L}^{(free)}_{(e)} (Higgs) =   \frac{1}{2}~
\partial_\mu
\left( \begin{array}{cccc} H_D (1)  \\ \\
H_D (2)  \\
 \end{array} \right)^\dag \cdot \partial^\mu
\left( \begin{array}{cccc} H_D (1)  \\ \\
  H_D (2) \\
 \end{array} \right) - \frac{1}{2}
 ~ M^2_D (H) \left( \begin{array}{cccc} H_D (1)  \\ \\
H_D (2)  \\
 \end{array} \right)^\dag \cdot \left( \begin{array}{cccc} H_D (1)  \\ \\
  H_D (2) \\
 \end{array} \right) \nonumber \\ \nonumber \\
&& H_D (1) = \frac{\sigma_{(1)} (x)}{2}: ~ H_D (2)  =
\frac{\sigma_{(2)} (x)}{2},
\end{eqnarray}
such that the doublet mass $M_D (H)$ is:

\begin{eqnarray}
&& M_D (H)= M_W \sqrt{ \frac{20}{3} + \frac{4 g_2^2}{3 g^2_1 }}~
 \thickapprox 213.7 GeV ;
\end{eqnarray}

c) All  remaining cases correspond to the singlet states $H_S (i)$
of the Higgs boson. Both the mass of the states and  explicit form
of them depend strongly on the gauge of the initial gauge fields
${\cal A}^a_\mu (x)$ and ${\cal B}^a_\mu (x)$. The Lagrangian
governing such states is:

\begin{eqnarray}
&& {\cal L}^{(free)}_{(e)} (Higgs) =  \frac{1}{2}(\partial_\mu H_S
(i) )
 (\partial^\mu H_S (i) ) - \frac{1}{2} M^2_S (H) ~H^2_S (i) ,
\end{eqnarray}
where $H_S (i)$ is the field of the Higgs mode which is directly
expressed via the functions $\sigma_{(i)} (x)$.

When $\sigma_{(1 )} (x) = \sigma_{(3)} (x) = 0 ;~ \sigma_{(2)}
(x)\ne 0 $, or $\sigma_{(2)} (x) = \sigma_{(3)} (x) = 0 ;  ~
\sigma_{(1)} (x)\ne 0 $, or  $\sigma_{(1 )} (x) = \sigma_{(2)} (x)
\ne 0 ;  \sigma_{(3)} (x)=0 $, the Higgs mass is

\begin{eqnarray}
M_S (H) = M_W \sqrt{ \frac{20}{3} + \frac{4 g_2^2}{3 g^2_1 }}~
 \thickapprox 213.7 GeV.
 \end{eqnarray}

In the case $\sigma_{(1 )} (x) = \sigma_{(2)} (x) = 0 ;~~
\sigma_{(3)} (x)\ne 0 $ the boson mass is

\begin{eqnarray}
M_H (S) = M_W~ \sqrt{\frac{4}{3}}= 92.8  GeV,
\end{eqnarray}

Provided that    $\sigma \equiv \sigma_{(1)} (x) = \sigma_{(2)}
(x) = \sigma_{(3)} (x)\ne 0 $,   the singlet mass is equal to

\begin{eqnarray}
M_S (H) = M_W \sqrt{ \frac{44}{9} + \frac{8 g_2^2}{9 g^2_1 }}~
 \thickapprox 182.6 GeV .
\end{eqnarray}

When    $\sigma \equiv \sigma_{(1)} (x) = \sigma_{(3)} (x)  \ne 0
; \sigma_{(2)} (x) = 0 $, or $\sigma \equiv \sigma_{(2)} (x) =
\sigma_{(3)} (x) \ne 0 ; \sigma_{(1)} (x) = 0 $,  the  mass is

\begin{eqnarray}
M_S (H) = M_W \sqrt{ 4 + \frac{2 g_2^2}{3 g^2_1 }}~
 \thickapprox 164.8 GeV .
\end{eqnarray}

The situations   $\sigma_{(2)} (x) \ne \sigma_{(1)} (x) =
\sigma_{(3)} (x) \ne 0; \sigma_{(2)} (x)\ne 0 $, and $\sigma_{(1)}
(x) \ne \sigma_{(2)} (x) = \sigma_{(3)} (x) \ne 0; \sigma_{(1)}
(x)\ne 0 $ lead to arising two singlets which masses are given by
the formulae (31), (33). Two singlets also arise when
$\sigma_{(3)} (x) \ne \sigma_{(1)} (x) = \sigma_{(2)} (x) \ne 0;
\sigma_{(3)} (x)\ne 0 $. In such case the Higgs mass is given by
Eqs.(31), (32).

Finally, it is obviously possible the situation when $\sigma_{(1)}
(x) = \sigma_{(2)} (x) = \sigma_{(3)} (x) = 0 $. It means that the
Higgs boson mode appears to be  unexcited in such case.

Thus, there is the spectrum of the mass states of the Higgs bosons
which also include  the situation when the Higgs  degrees of
freedom appear to be unexcited. Since the source of the Higgs
bosons in the developed model is gauge fields, what case  takes
place can be only experimentally revealed. In this way, as soon as
any  gauge of the  fields ${\cal A}^a_\nu (x)$ and ${\cal B}^a_\nu
(x)$ is realizable the different mass states can generally arise
in different experiments that can create additional problems in
Higgs identification.

\section { Conclusion}

The unified theory  of the electromagnetic and weak interactions
is developed in the paper. On a basis of violation of the initial
$U_L(2)\bigotimes U_R(2) $ gauge symmetry, and in the absence of a
priory self-interaction scalar field the $SU_L(2)\bigotimes U_R(1)
$ Lagrangian is derived. The derived Lagrangian is found to take
into account correctly interactions between particles via  both
the charged and neutral currents as it takes place in the
SM\cite{4,5,6}. In this way, all massive boson field (including
the Higgs boson),  massive fermion fields as well as
electromagnetic field naturally arise as the superposition of the
deferent modes of the initial  gauge fields. On a basis of the
experimental data for the fine structure coupling constant, Fermi
constant and $W$ boson mass, the coupling constant $g_1$ and $g_2$
of the developed consideration  as well as the mass of the
$Z^{(0)}$ boson are calculated. The structure of the Higgs field
is studied in detail. It is shown that the different gauges of the
initial fields ${\cal A}^a_\nu (x)$ and ${\cal B}^a_\nu (x)$ lead
to the different states of the Higgs bosons which include the
situation when the Higgs degrees of freedom appear to be
unexcited. In the cases when the Higgs mode arises  the masse of
the Higgs boson is found to be in the interval $M_H = 92.8 \div
231.7 ~GeV$.

\end{document}